\documentclass[doublecol]{epl2}
\usepackage{graphicx}

\title{Correlation functions for the detection of Wigner molecules in a one-channel Luttinger liquid quantum dot}
\shorttitle{Correlation functions for the detection of Wigner molecules\ldots} 

\author{F. M. Gambetta\inst{1} \and N. Traverso Ziani\inst{2} \and F. Cavaliere\inst{1,2} \and M. Sassetti\inst{1,2}}
\shortauthor{N. Traverso Ziani \etal}

\institute{
  \inst{1}Dipartimento di Fisica, Universit\`a di Genova, Via Dodecaneso 33,
  16146, Genova, Italy.  \\
  \inst{2}CNR-SPIN, Via Dodecaneso 33,
  16146, Genova, Italy.\\}
\abstract{In one-channel, finite-size Luttinger one-dimensional quantum dots, both Friedel oscillations and Wigner correlations induce oscillations in the electron density with the {\em same} wavelength, pinned at the same position. Therefore, observing such a property does not provide any hint about the formation of a Wigner molecule when electrons interact strongly and other tools must be employed to assess the formation of such correlated states. We compare here the behavior of three different correlation functions and demonstrate that the {\em integrated two point correlation function}, which represents the probability density of finding two particles at a given distance, is the only faithful estimator for the formation of a correlated Wigner molecule.}
\begin{document}

\maketitle
\section{Introduction}
\noindent Strongly interacting one-dimensional (1D) systems are attracting an increasing attention: 1D quantum dots
in carbon nanotubes~\cite{nat1,nat2}, semiconducting heterostructures and nano-wires~\cite{qw2,qw3}, helical 1D edge states in two dimensional topological insulators~\cite{ti1,ti2,ti3,ti4} and in spin-orbit coupled quantum wires~\cite{quoy,meng} and the cold atom simulators~\cite{lewenstein},
are only some of the most prominent examples.\\
From the experimental point of view, 1D quantum dots~\cite{koudots} represent an invaluable tool:
their small dimension reinforce interaction effects, allowing for their detection. Moreover with transport spectroscopy one can investigate the low energy properties~\cite{koudots},
and, in combination with local probes, obtain information such as the electron and spin density~\cite{linear,halperin,afm,stm,topologicalprb,zhukov,zhukov2}.\\
From the theoretical perspective, 1D interacting electrons have been intensely studied:
aside from numerical techniques, such as exact diagonalization~\cite{secchi1,secchi2},
quantum Montecarlo~\cite{russi,pimc1}, density functional theory~\cite{ba1,ba2,ba3,polinirontani}, and density matrix
renormalization group~\cite{dmrg},
exact solutions are possible thanks to the Bethe ansatz solutions~\cite{bethe,libronelieb}. Moreover a powerful field theory,
the Luttinger liquid theory~\cite{haldane,giamarchi,voit}, is also available.
Its validity ranges from semiconducting quantum wires~\cite{wireslutt}, carbon nanotubes~\cite{cntgogolin,cntgrifoni,yoshioka,univ},
edges in the integer and fractional quantum Hall effect~\cite{hall1,hall2,hall3}, to two dimensional topological insulators~\cite{ti2,ti3}, spin-orbit coupled quantum wires~\cite{quoy,meng}, cold atoms in 1D optical lattices~\cite{ca1,bosons1},
spin systems~\cite{spin}, and mesoscopic circuits~\cite{mesosafi}. The main feature of the Luttinger liquid is the bosonic character of its
low energy excitations~\cite{haldane}, even for strong interactions. This issue allows for analytical calculations in any interaction regime.\\
{\noindent In order to describe {\em finite-size}, sharply confined systems, such as quantum dots, one has to to adopt {\em finite-size} boundary conditions (FBC), such as open-boundary conditions, which strongly differ from periodic boundary conditions (PBC)~\cite{giamarchi,open,yoshioka}. In general, the effect of FBC is to halve the number of channels of a Luttinger liquid theory with respect to the case of PBC. For example, finite-size spinful 1D Luttinger liquids - which for PBC have four independent channels, exhibit only the two charge and spin channels, as reflections at the boundaries mixes left- and right-movers~\cite{giamarchi}. This system will be dubbed henceforth a {\em two-channel Luttinger liquids} (2LL). Similarly, systems sporting two independent channels with PBC will be enforced, by FBC, to become one-channel Luttinger liquids (1LL). A typical and very relevant example of this are the helical LL (HLL) occurring at the edge of topological insulators~\cite{ti1,ti2,ti3,ti4}. Here, spin-momentum locking pairs the up (down) spin direction to right (left) movers yielding a 2LL with PBC. The additional presence of ferromagnetic barriers~\cite{timm,topologicalprb,vish} enforces twisted boundary conditions which mix spin and chirality of electrons~\cite{timm,topologicalprb}, resulting in a 1LL. Other notable examples of 1LLs are fully spin-polarized 2LLs and the spin-incoherent Luttinger liquid (SILL)~\cite{fiete1,fiete2,mfg2,mfg}. The former can be easily achieved by applying a magnetic field to a 2LL. The SILL describes strongly interacting spinful electrons in the temperature regime $D_\sigma\ll k_BT \ll D_\rho$, $D_\sigma$ and $D_\rho$ being the bandwidth of spin and charge excitations~\cite{matveev}. The corresponding Hamiltonian maps onto that of a 1LL describing spinless fermions called holons~\cite{mfg2,mfg}.\\
\noindent Therefore, the 1LL is far from being a pure theoretical model but represents nowadays a very interesting and lively subject of investigation.\\

\noindent The most striking effect of strong, long range interactions in 1D quantum dots is the formation of the Wigner molecule~\cite{wigner,wigmol1,wigmol2,vignale,schulz}, which is the finite-size counterpart of a Wigner crystal. Indeed, due to fluctuations, no Wigner {\em crystal} can occur in 1D. However, when the electronic correlation length exceeds the length of the sample a correlated molecular state arises~\cite{wigmol1,wigmol2}. In a semiclassical picture, one can picture such a state as a regular array of $N$ electrons, each of which free to oscillate around its equilibrium position~\cite{giamarchi}.\\
\noindent In a 2LL, the density profile shows a competition between Friedel oscillations, a finite size effect present even in the absence of interactions, and Wigner oscillations, a pure interaction effect~\cite{schulz,afm,dmrg,ba1,mantelli}. The two oscillations are easily distinguishable on the basis of their different wavevectors, with Friedel (Wigner) oscillations being characterized by the wavevector $k_F=2k_0$ ($k_W=4k_0$), with $k_0$ the Fermi momentum~\cite{afm}. It has been demonstrated that Friedel oscillations dominate for weak interactions, while Wigner oscillations dominate for strong interactions~\cite{giamarchi,voit}.
However, while the presence of Wigner oscillations is a clear signature of interaction effects, the degree of correlation cannot be inspected within the electron density alone~\cite{polinirontani}. It has in fact been shown that in the presence of zero range interaction, Wigner oscillations appear in the electron density, while correlation functions unveil the uncorrelated nature of the state~\cite{polinirontani}.\\
\noindent In 1LL quantum dots on the other side both Friedel and Wigner oscillations have the {\em same} wavevector ($2k_0$)~\cite{fiete1,fiete2,sablikov}. While in the noninteracting case it is easy to show, using the wavefunctions for a particle in a box, that the number of peaks of the local electron density of the system containing $N$ particles is precisely $N$, in the case of strong interactions the $N$ electrons of the Wigner molecule, oscillating around their equilibrium position, give rise again to $N$ distinct peaks in the density. As a consequence it is not even possible to discriminate between finite size and interaction effects by studying the electron density alone~\cite{sablikov,safi}. The interplay between finite size and interaction effects and correlations must therefore be clarified in more details.\\}

\noindent The aim of this letter is to detect the transition from an interacting, liquid-like state to a strongly correlated Wigner molecule in a 1LL. We identify a suitable tool to confirm the presence of a Wigner molecule in a 1LL, comparing three different correlation functions and demonstrate that the probability density of finding two particles at a given distance is the best tool to detect Wigner correlations. By means of this, we show that strong interactions induce a crossover from an interacting but still liquid-like state to a Wigner molecule, in analogy to the case of a 2LL. We estimate the threshold value of the interaction parameter for the appearance of the Wigner molecule as $g\sim 0.45$.\\

\noindent The outline of the Letter is the following:\\
\noindent In order to elucidate the differences between 2LL and 1LL, we start briefly reviewing the properties of the former, recalling the crossover between Friedel and Wigner oscillations in the electron density. Subsequently, we turn to the case of a 1LL where we show that the electron density displays no peculiar crossover when the interaction strength is increased. We then introduce three possible tools to detect the emergence of correlations and show that studying the conditioned probability to find two electrons at a given distance is {among} the most sensitive and reliable ways to assess Wigner correlations.
\section{Two-channel Luttinger liquid}
We start by recalling the main results for a 2LL.
The Hamiltonian $H_{2LL}$ reads ($\hbar=1$)~\cite{open}
\begin{equation}
H_{2LL}=H_N+H_{b},\label{eq:toth}
\end{equation}
with
\begin{eqnarray}
H_N&=&\frac{E_{\rho}}{2}N^2+\frac{E_{\sigma}}{2}N_{\sigma}^2, \label{eq:hn}\\
H_{b}&=&\sum_{n=1}^{\infty}\left[\varepsilon_{\rho}nd^\dag_{\rho,n}d_{\rho,n}+\varepsilon_{\sigma}nd^\dag_{\sigma,n}d_{\sigma,n}\right]\, . \label{eq:hp}
\end{eqnarray}
Here, $N=N_++N_-$, $N_\sigma=N_+-N_-$, $N_{s}$ (with $s=\pm$) is the number of electrons with spin projection up/down and $H_{N}$ represents the contribution of the zero modes, with $E_{\nu}=\pi v_{\nu}/2Lg_{\nu}$ written in terms of the velocity $v_{\nu}$ of the mode
$\nu=\rho,\sigma$, of the system length $L$ and of the Luttinger parameters $g_{\nu}$. For
repulsive interactions one has $g_{\rho}=g<1$, while $g=1$ corresponds
to the noninteracting limit. On the other hand, $g_{\sigma}=1$ for an
SU(2) invariant theory. The velocity $v_{\rho}=v_{0}/g$ of the charged mode (with $v_{0}$ the Fermi velocity)
is renormalized by the interactions while the spin modes velocity is $v_{\sigma}=v_{0}$.\\

\noindent The term $H_{b}$ describes collective, quantized charge and
spin density waves with boson operators $d_{\nu,n}$ and energy $\varepsilon_{\nu}=\pi v_{\nu}/L$.\\
\noindent The electron field operator $\Psi_s(x)$ satisfying open
boundary conditions $\Psi_s(0)=\Psi_s(L)=0$ is~\cite{open} $\Psi_s(x)=\psi_{s,+}(x)-\psi_{s,+}(-x)$,
where $\psi_{s,+}(x)$ is a $2L$-periodic fermion field that admits the bosonic representation~\cite{open}
\begin{equation}
\psi_{s,+}(x)=\frac{\eta_s}{\sqrt{2\pi\alpha}}e^{-i\theta_{s}}
\,e^{i\frac{\pi  N_sx}{L}}e^{i\frac{\Phi_{\rho}(x)+s\Phi_\sigma(x)}{\sqrt{2}}}\,. \label{eq:opright}
\end{equation}
Here, $\alpha$ is the cutoff length, set as $\alpha=L/(\pi N)$, $\theta_{s}$
satisfies $[\theta_{s}, N_{s'}]=i\delta_{s,s'}$, and $\eta_s$ fulfill
$ \eta_s\eta_{s'}+\eta_{s'}\eta_s=2\delta_{s,s'}$, allowing the right
anticommutation relations for different spins. The boson fields
$\Phi_{\rho}(x)$, $\Phi_{\sigma}(x)$ are given by
\begin{equation}
\!\!\!\!\Phi_{\nu}(x)\!=\!\sum_{n=1}^{\infty}\frac{e^{-\frac{n\pi\alpha}{L}}}{\sqrt{g_{\nu}n}}
\left[\!\left(\cos{\frac{n\pi x}{L}}\!-\!ig_{\nu}\sin{\frac{n\pi x}{L}}\right)d^\dag_{\nu,n}\!+\mathrm{h.c.}\right].
\end{equation}
The particle density operator is $\rho_{2LL}(x)=\sum_{s=\pm}\rho_s(x)$ with $\rho_{s}(x)=\Psi_{s}^{\dagger}(x)\Psi_{s}(x)$.
Following a standard procedure it can be bosonized~\cite{haldane,schulz,safi,mantelli,footnote}
\begin{equation}
\!\!\!\rho_{2LL}(x)\!=\!\frac{N}{L} + \frac{\sqrt{2}}{\pi}\partial_x \varphi_\rho(x)\!+\!F\sum_{s=\pm}\rho_s^F(x)\!+\!(1\!-\!F){\rho^W(x)}.
\end{equation}
Here
\begin{eqnarray}
\rho_s^F(x)&=&-\frac{N_s}{L}\cos\left[\frac{2N_s\pi x}{L}-2f(x)-2\varphi_{s}(x)\right]\, ,\\
\rho^W\!(x)\!&=&\!-\frac{ N}{L}\cos\!\left[\frac{2N\pi}{L}-4f(x)-2\sqrt{2}\varphi_\rho(x)\right]\, ,\\
\varphi_s(x)&=&\frac{\varphi_\rho(x)+s\varphi_\sigma(x)}{\sqrt{2}},\\
\varphi_{\rho/\sigma}(x)&=&\frac{1}{2}\left[
  \Phi_{\rho/\sigma}(-x)-\Phi_{\rho/\sigma}(x)\right]\,,\label{eq:varphi}\\
  f(x)&=&\frac{1}{2}\tan^{-1}\left(\frac{\sin(2\pi x/L)}{e^{\pi\alpha/L}-\cos (2\pi x/L)}\right)\, ,\label{eq:fofx}
\end{eqnarray}
where $0\leq F\leq 1$ models the relative weight of the Friedel and the Wigner contributions to the density~\cite{dmrg}.\\
\noindent From the above equations one can see that Friedel oscillations display
$N/2$ peaks for even $N$ and $N/2+1/2$ for odd $N$), while Wigner oscillations always display $N$ peaks~\cite{mantelli}.
\begin{figure}[htbp]
\begin{center}
\includegraphics[width=7.2cm,keepaspectratio]{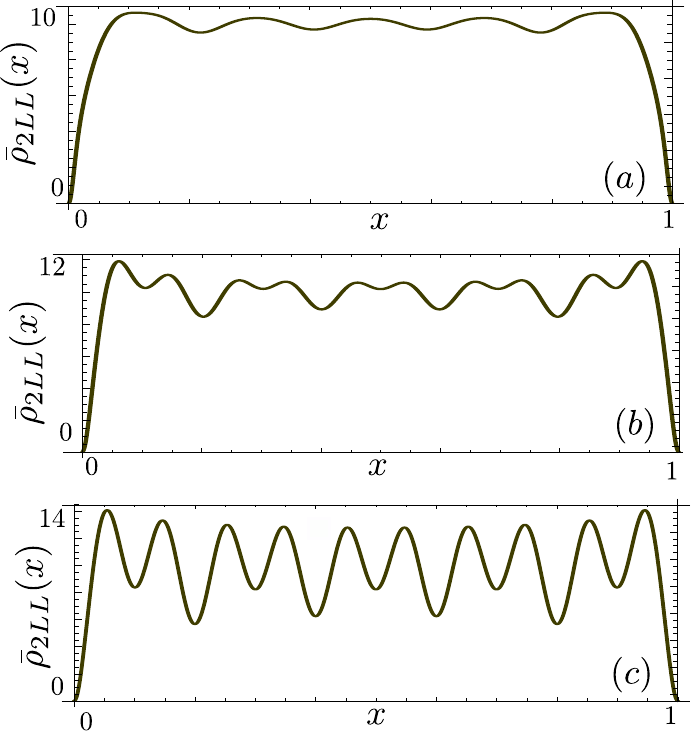}
\caption{Plot of $\bar{\rho}_{2LL}(x)$ (units $1/L$) as a function of $x$ (units $L$) for $N=10$
and different interaction strengths: (a) $g=1$; (b) $g=0.4$; (c) $g=0.1$. In all panels $F=0.5$, $\alpha=L/(10\pi)$, $T=0$.}
\label{fig:fig1}
\end{center}
\end{figure}
Figure~\ref{fig:fig1} shows the zero-temperature averaged electron density $\bar{\rho}_{2LL}(x)=\langle N,\bar{N}_{\sigma}|\rho_{2LL}(x)|N,\bar{N}_{\sigma}\rangle$ for different interactions. Here $|N,\bar{N}_{\sigma}\rangle$ is the ground state for $N$ electrons and $\bar{N}_{\sigma}=0$ ($\bar{N}_{\sigma}=\pm 1$) for even (odd) $N$. The average can be computed analytically~\cite{afm}: results are not quoted here for simplicity. While in the non interacting case ($g=1$) only Friedel oscillations appear, Fig.~\ref{fig:fig1}(a), increasing interactions a competition between Friedel and Wigner oscillations develops as shown in Fig.~\ref{fig:fig1}(b). Finally, in the strong interaction regime ($g\to0$) well developed Wigner oscillations are present, see Fig.~\ref{fig:fig1}(c). Indeed, it can be analytically proven (not shown) that Wigner oscillations dominate over Friedel ones in the density for $g\leq 1/3$~\cite{safi}.\\
\noindent We can thus conclude that, even though the presence of a strongly correlated Wigner molecule must be inferred by higher order correlation functions~\cite{polinirontani}, the crossover between the weak and the strong interaction regimes can be inspected in terms of oscillations of the electron density.
\section{One-channel Luttinger liquids}
\begin{figure}[htbp]
\begin{center}
\includegraphics[width=7.2cm,keepaspectratio]{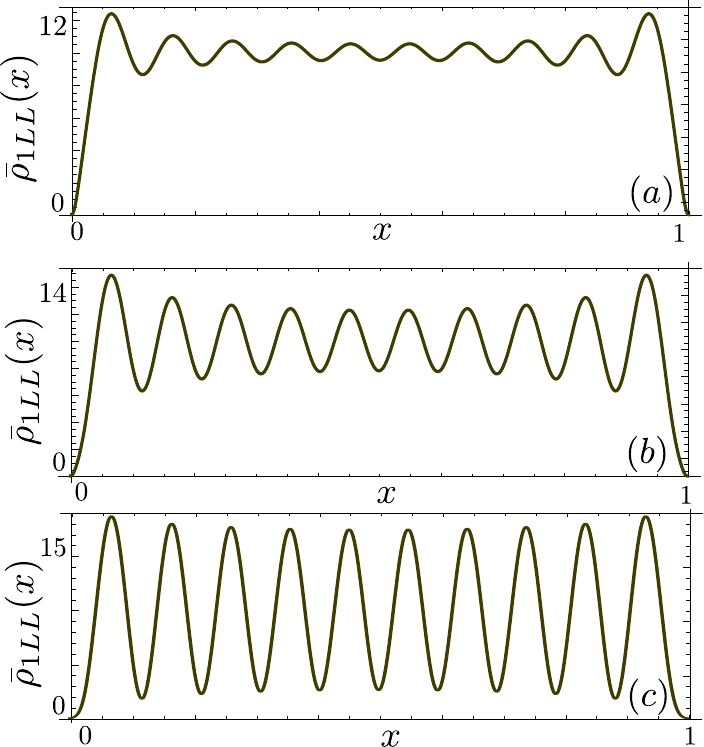}
\caption{Plot of $\bar{\rho}_{1LL}(x)$ (units $1/L$) as a function of $x$ (units $L$) for $N=10$
and different interaction strengths: (a) $g=1$, (b) $g=0.5$, (c) $g=0.1$. In both panels $\alpha=L/(10\pi)$, $T=0$.}
\label{fig:fig2}
\end{center}
\end{figure}

Let us now consider a 1LL. {For definiteness, we will consider the case of a fully spin-polarized 2LL although all the results presented here apply to the other cases of 1LL described in the introduction.} The Hamiltonian is~\cite{haldane,voit,giamarchi}
\begin{equation}
H_{1LL}=\frac{E_{0}}{2}N^2+\varepsilon_{0}\sum_{n=1}^{+\infty}n b^\dag_n b_n,
\end{equation}
with $E_{0}=\pi v_{0}/Lg^2$ and $\varepsilon_{0}=\pi v_{0}/Lg$.
The electron operator $\Psi(x)$, satisfying $\Psi(0)=\Psi(L)=0$, is~\cite{sablikov,open}
$\Psi(x)=\psi_R(x)-\psi_R(-x)$, with the $2L$-periodic field $\psi_R(x)$ given by
\begin{equation}
\psi_R(x)=\frac{1}{\sqrt{2\pi\alpha}}e^{-i\theta}e^{i\frac{\pi Nx}{L}}e^{i\Phi(x)}.
\end{equation}
Here the bosonic field $\Phi(x)$ is
\begin{equation}
\!\!\Phi(x)\!=\!\sum_{n>0}\frac{e^{-\frac{\alpha\pi n}{L}}}{\sqrt{gn}}
\!\left[\!\cos\!\left(\!\frac{n\pi x}{L}\!\right)\!-\!i{g}\sin{\!\left(\!\frac{n\pi x}{L}\!\right)}\!\right]b_n+{\mathrm h.c.}\ ,
\end{equation}
with $[\theta, N]=i$. The electron density operator $\rho_{1LL}(x)=\Psi^{\dag}(x)\Psi(x)$ can be bosonized as~\cite{schulz,mantelli,afm,safi,footnote}
\begin{equation}
\rho_{1LL}(x)=\frac{N}{L}-\frac{\partial_x\varphi(x)}{2\pi}-\frac{N}{L}\cos\!\left[\frac{2\pi N x}{L}-2\varphi(x)-2 f(x)\right],
\end{equation}
with $f(x)$ defined as in Eq.~(\ref{eq:fofx}) and
\begin{equation}
\varphi(x)=i\sqrt{g}\sum_{n>0}\frac{e^{-\frac{n\pi\alpha}{2L}}}{\sqrt{n}}\sin \left(\frac{n\pi x}{L}\right)\left(b^\dag_{n}-b_{n}\right).\\
\end{equation}
The zero-temperature average electron density $\bar{\rho}_{1LL}(x)=\langle N|\rho_{1LL}(x)|N\rangle$ (with $|N\rangle$ the ground state of the 1LL with $N$ electrons) is
\begin{eqnarray}
\bar{\rho}_{1LL}(x)&=&\frac{N}{L}-\frac{N}{L}\cos\left[\frac{2N\pi x}{L}\!-\!2f(x)\right]K(x)\\
K(x)&=&\left[\frac{\sinh\left(\frac{\pi\alpha}{L}\right)}{\sqrt{\sinh^2\left(\frac{\pi\alpha}{L}\right)+\sin^2\left(\frac{\pi x}{L}\right)}}\right]^g
\end{eqnarray}
and is shown in Fig.~\ref{fig:fig2} for different interactions.\\
\noindent Even in the noninteracting case $g=1$, Fig.~\ref{fig:fig2}(a), the electron density exhibits $N$ peaks. Increasing interactions only results in a continuous enhancement of the peak-to-valley ratio, Figs.~\ref{fig:fig2}(b,c). Indeed, both the finite-size Friedel oscillations and Wigner oscillations have the same wavelength $\sim L/N$. Studying the density alone, it is therefore impossible to assess if a critical value $g_{c}(N)$ exists such that for $g<g_{c}(N)$ interactions overcome finite size effects.\\

\noindent To detect the crossover towards a strongly interacting regime and finally to the formation of a strongly correlated Wigner molecule, which is the
 task of this Letter, we now compare three different correlation functions, namely\\

\noindent($i$) the density-density correlation function;\\
\noindent($ii$) the pair correlation function;\\
\noindent($iii$) the probability density of finding two electrons at distance $x$.\\

\noindent($i$) The density-density correlation function is
\begin{equation}
d_{1LL}(N,x,y)=\langle N|\rho_{1LL}(x)\rho_{1LL}(y)|N\rangle\, .
\end{equation}
Although its calculation can be carried out analytically, the explicit result is omitted for simplicity.
\begin{figure}[htbp]
\begin{center}
\includegraphics[width=7.2cm,keepaspectratio]{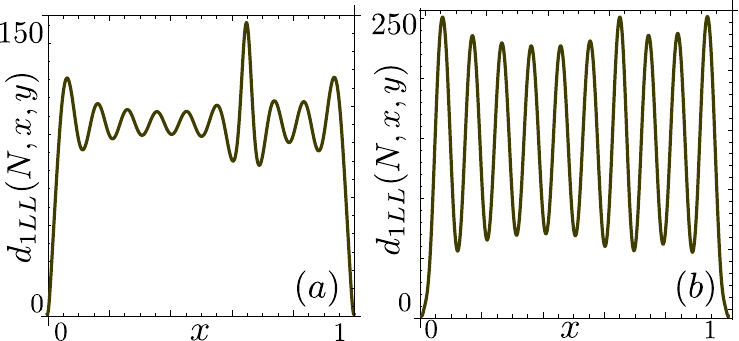}
\caption{Plot of $d_{1LL}(N,x,y)$ (units $1/L^2$) as a function of $x$ (units $L$)
for different interaction strength (a) $g=1$; (b) $g=0.2$. In both panels $N=10$, $y=0.65 L$, and $\alpha=L/(10\pi)$, $T=0$.}
\label{fig:fig3}
\end{center}
\end{figure}
A plot of $d_{1LL}(N,x,y)$ is shown in Fig.~\ref{fig:fig3}. We chose for $y$ the location of one of the maxima in the electron density, in order to maximize the contrast. The function exhibits $N$ distinct peaks, the auto-correlation one at $x=y$ being dominant, especially for the non-interacting case shown in Fig.~\ref{fig:fig3}(a). In the strongly interacting regime the only noticeable effect is an increase of the peak-to-valley ratio as seen in Fig.~\ref{fig:fig3}(b). Thus, this quantity behaves qualitatively in the same way as the electron density and no clear onset of an interaction-induced crossover towards the correlated Wigner regime for can be detected.\\

\noindent($ii$) The pair correlation function $g_{1LL}(N,x,y)$ is~\cite{vignale}
\begin{equation}
g_{1LL}(N,x,y)=\frac{\langle N|\Psi^\dag(x)\Psi^\dag(y)\Psi(y)\Psi(x)|N\rangle}{\bar{\rho}_{1LL}(x)\bar{\rho}_{1LL}(y)}
\end{equation}
and has often been employed for characterizing strongly correlated systems~\cite{vignale}.
\begin{figure}[htbp]
\begin{center}
\includegraphics[width=7.2cm,keepaspectratio]{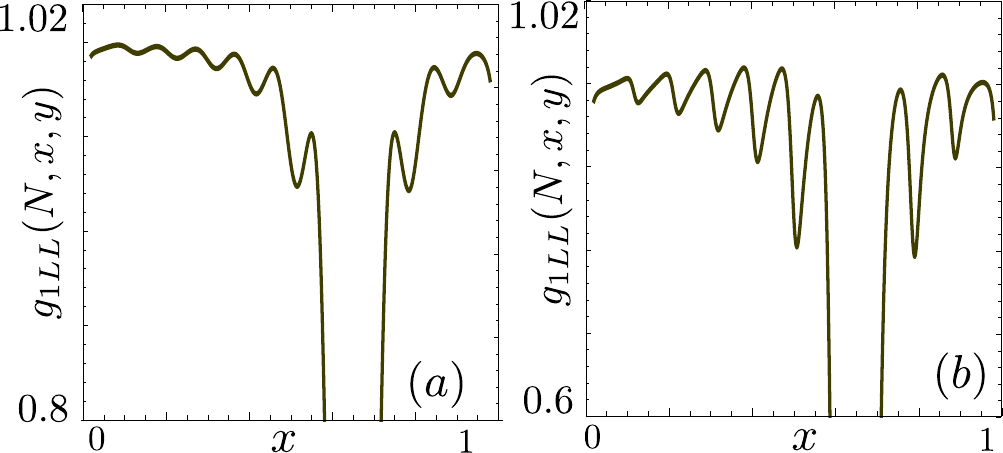}
\caption{Plot of $g_{1LL}(N,x,y)$ as a function of $x$ (units $L$)
for different interaction strengths (a) $g=1$; (b) $g=0.2$. In both panels $N=10$, $y=0.65 L$, and $\alpha=L/(10\pi)$, $T=0$.}
\label{fig:fig4}
\end{center}
\end{figure}
A plot of {$g_{1LL}(N,x,y)$} for $N=10$ and $y$ located at a maximum of the electron density is shown in Fig.~\ref{fig:fig4}. The most notable feature is the presence of a Pauli hole, namely the sharp collapse for $|x-y|\ll L/N$ due to the Pauli exclusion principle. For enhancing the visibility of the oscillations we set the scale to a restricted region, the Pauli hole being unimportant for our aims. Both the noninteracting case (Fig.~\ref{fig:fig4}(a)) and the strongly interacting (Fig.~\ref{fig:fig4}(b)) have the same number of peaks, namely $N-1$. Again, the effect of interactions is to enhance the peak-to-valley ratio with no additional feature emerging. We can therefore rule out also this quantity as a clear-cut estimator of the emergence of Wigner correlations in a 1LL.\\

\noindent($iii$) The probability density $P_{1LL}(x)$ of finding two electrons at distance $x$ has been recently proposed as a tool to detect Wigner correlations in 1D~\cite{polinirontani}. It is defined as
\begin{equation}
P_{1LL}(N,x)=\frac{\int_{-\infty}^\infty dy \langle N|h(x,y)|N\rangle}{N(N-1)},
\end{equation}
where
\begin{equation}
h(x,y)=\Psi^\dag\left(y+\frac{x}{2}\right)\Psi^\dag\left(y-\frac{x}{2}\right)\Psi\left(y-\frac{x}{2}\right)\Psi\left(y+\frac{x}{2}\right).\nonumber
\end{equation}
While the quantum average can be analytically performed, the integration is carried out numerically.
\begin{figure}[htbp]
\begin{center}
\includegraphics[width=7.2cm,keepaspectratio]{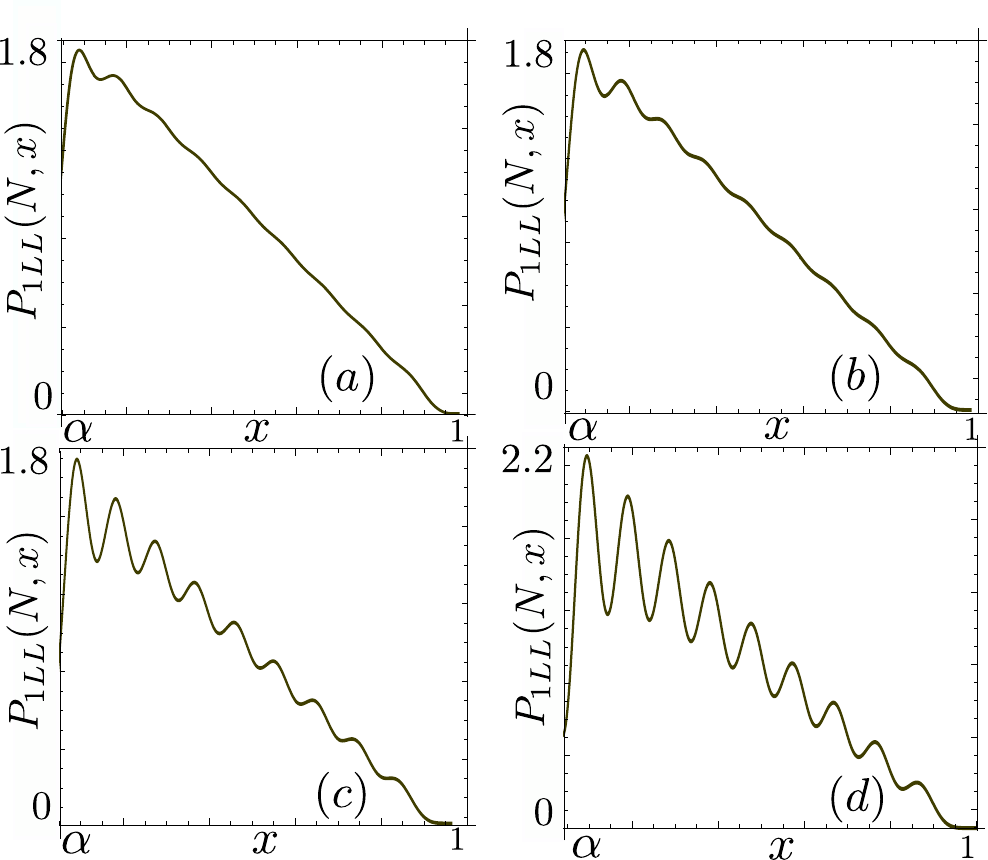}
\caption{Plot of $P_{1LL}(N,x)$ (units $1/L$) as a function of $x$ (units $L$)
and different interaction strength (a) $g=1$; (b) $g=0.8$; (c) $g=0.45$; (d) $g=0.25$. In all panels $N=10$ and $\alpha=L/(10\pi)$, $T=0$.}
\label{fig:fig5}
\end{center}
\end{figure}
Results for $N=10$ are shown in Fig.~\ref{fig:fig5} for $x>\alpha$ (for distances shorter than the cutoff $\alpha$ the calculation is not reliable). For zero interactions ($g=1$), Fig.~\ref{fig:fig5}(a), $P_{1LL}(N,x)$ exhibits a Pauli hole (whose precise form cannot be evaluated with our approach for $x<\alpha$) but is otherwise almost featureless with only a hint oscillations for $x\ll 1$. The overall decrease of $P_{1LL}(N,x)$ as $x$ is increased is due to the reduction of phase space in the integration. The absence of correlations in $P_{1LL}(N,x)$ confirms the liquid-like structure of the dot state for the noninteracting case. Increasing interactions to mild values, for $g=0.8$ shown in Fig.~\ref{fig:fig5}(b), signatures of correlations begin to develop with $P_{1LL}(N,x)$ showing three distinct maxima for $x<1/2$. Notice that due to the Pauli hole, the maximum number of maxima expected for $P_{1LL}(N,x)$ is $N-1$. This suggests to define $g_{c}(N)$ such that, for $g<g_{c}(N)$ one has $N-1$ distinct maxima in $P_{1LL}(N,x)$. We have performed extensive numerical scans of $P_{1LL}(N,x)$ for $5\leq N\leq30$ and found that the crossover is almost insensitive to the number of particles $0.45\leq g_{c}(N)<0.5$. Indeed, Figs.~\ref{fig:fig5}(c,d) show $P_{1LL}(N,x)$ for $g<g_{c}(10)\sim0.46$. Clearly, 9 distinct peaks located at $x=x_{n}\sim nL/10$ ($1\leq n\leq 9$) are present, which confirm the physical picture of an ordered molecular state.\\
{\noindent The origin of the better performance of $P_{1LL}(N,x)$ with respect to $g_{1LL}(N,x,y)$ and $d_{1LL}(N,x,y)$ lies in the integration over the coordinate of the reference position with respect to which the distance $x$ between correlated electrons is measured. Indeed, both in $g_{1LL}(N,x,y)$ and in $d_{1LL}(N,x,y)$ an unavoidable background of the uncorrelated density oscillations remains even in the weakly interacting regime. Such a background is washed by the integration and genuine electronic correlations emerge.}
\section{Conclusions}
\label{sec:conclusions} We have studied the formation of a correlated Wigner molecule in a one-dimensional finite-size system of $N$ electrons, described by a one-channel Luttinger liquid. Both Friedel and Wigner oscillations of the electron density are characterized by $N$ distinct peaks located at the same positions. Thus, the density does not bring information on the crossover from finite size to correlation effects as interactions increase. We have compared three different tools to investigate correlations among the electrons: the density-density correlation function, the pair correlation function, and the density probability of finding two electrons at a given distance. We have shown that the latter is the best detector of the onset of Wigner crystallization among the three considered in this paper. In a large range of particle numbers $5\leq N\leq 30$, the critical value of the Luttinger parameter $g_{c}(N)$ below which the correlations appears is in the range $0.45\leq g_{c}(N)<0.5$.
\acknowledgments
Financial support by MIUR via MIUR-FIRB2012, Grant No. RBFR1236VV is gratefully acknowledged.


\begin{thebibliography}{99}
\bibitem{footnote} Slightly different forms of the electron density can be employed~\cite{sablikov}. They do not lead to qualitative modifications of the results presented here.
\end{thebibliography}
\end{document}